\documentclass{PoS}
\usepackage[square,numbers]{natbib}
\usepackage{graphicx}
\usepackage{amsmath}
\usepackage{amssymb}
\usepackage{xfrac}
\usepackage{esvect}
\usepackage{varwidth} 
\newcommand{\A}[1]{\sqrt{1-3\cos^{2} #1}} 

\title{Modelling the polarisation signatures detected from the first white dwarf pulsar AR Sco.}

\ShortTitle{Polarisation modelling for the white dwarf pulsar AR Sco}

\author{\speaker{L du Plessis$^1$},  Z Wadiasingh$^{1,2}$,C Venter$^1$, A K Harding$^2$, S~Chandra$^1$, P J Meintjes$^3$ \\
        $^1$ Centre for Space Research, Private Bag X6001, North-West University, 
Potchefstroom, South Africa, 2520 \\
        $^2$ Astrophysics Science Division, Code 663 NASA's Goddard Space Flight Center, Greenbelt, MD 20771, U.S.A. \\
        $^3$ Department of Physics, University of the Free State, PO Box 339, Bloemfontein, South Africa, 9300 \\
        E-mail: \email{louisdp95@gmail.com}
        }

\abstract{ Marsh et al. detected radio and optical pulsations from the binary system AR Scorpii (AR Sco). This system, with an orbital period of 3.56 h, is composed of a cool, low-mass star and a white dwarf with a spin period of 1.95 min. Optical observations by Buckley et al. showed that the polarimetric emission from the white dwarf is strongly linearly polarised ( up to $\sim40\%$) with periodically changing intensities. This periodic non-thermal emission is thought to be powered by the highly magnetised ($ 5 \times 10^{8} $ G) white dwarf that is spinning down. The morphology of the polarisation signal, namely the position angle plotted against the phase angle, is similar to that seen in many radio pulsars. In this paper we demonstrate that we can fit the traditional pulsar rotating vector model to the optical position angle. We used a Markov-chain-Monte-Carlo technique to find the best fit for the model yielding a magnetic inclination angle of $\alpha = (86.6^{+3.0}_{-2.8})^{\circ}$ and an observer angle of $\zeta = (60.5^{+5.3}_{-6.1})^{\circ}$. This modelling supports the scenario that the synchtrotron emission originates above the polar caps of the white dwarf pulsar and that the latter is an orthogonal rotator.}

\FullConference{High Energy Astrophysics in Southern Africa - HEASA2018\\
		1-3 August, 2018\\
		Parys, Free State, South Africa}

\begin{document}

\section{Introduction}
\noindent  AR Sco is is a unique binary system, which has some similar properties to the propeller system, AE Aquarii \citep{2016Natur.537..374M}. Marsh et al. \cite{2016Natur.537..374M} observed pulsating emission ranging from the radio to the X-ray band. They found that the orbital period is 3.56 hours and the white dwarf pulsar's spin period is 1.95 minutes. A change in period $(\dot{P})$ of $3.9\times10^{-13}\rm{ss^{-1}}$ was observed \citep{2016Natur.537..374M}, but the value of $\dot{P}$ was later questioned by Potter and Buckley \citep{2018MNRAS.478L..78P}, where Stiller et al. \cite{Stiller2018} found a revised $\dot{P}$ of about twice the reported value from Marsh et al. \cite{2016Natur.537..374M}.

\noindent The aim of this paper is to show that the standard rotating vector model \citep{1969ApL.....3..225R} can be used to model the optical polarisation position angle of the white dwarf pulsar to constrain the origin of the emission. For more details, see Du Plessis et al., in preparation. 
\subsection{Observations}  
\noindent The light curve from the right column of Figure 2 in Marsh et al. \cite{2016Natur.537..374M} shows a double peak with a larger first peak followed by a smaller second peak, with a separation of $\sim180^{\circ}$ in phase. This could indicate that the magnetic inclination angle $\alpha$ is close to $90^{\circ}$ since the observation shows radiation from both poles. Buckley et al. \cite{2017NatAs...1E..29B} found the emission of AR Sco was strongly linearly polarized (up to $\sim 40\%$) and observed a $180^{\circ}$ swing in the polarisation position angle (PPA). Given the lack of evidence for mass loss from the secondary or accretion \citep{2016Natur.537..374M}, Buckley et al. \cite{2017NatAs...1E..29B} interpreted the spin-down power loss as due to magnetic dipole radiation. AR Sco is thus the first known white dwarf pulsar, since it is pulsating  with predominantly non-thermal emission,  analogous to pulsating neutron stars. 

\subsection{The Rotating Vector Model (RVM)}

\noindent Pulsars are known to be rotating neutron stars with two radio emission cones located at the magnetic poles. The inclination angle of the magnetic axis with respect to the rotation axis is represented by $\alpha$, the observer's line of sight by $\zeta$, and the impact angle $\beta =\zeta - \alpha$. Using the RVM, the PPA $\psi$ is given by
\begin{equation} \label{eq:1}
\tan(\psi-\psi_{0})=\frac{\sin\alpha\sin(\phi-\phi_{0})}{\sin\zeta\cos\alpha-\cos\zeta\sin\alpha\cos(\phi-\phi_{0})},
\end{equation}
with $\phi$ the sweeping (phase) angle.
\noindent The parameters $\phi_{0}$ and $\psi_{0}$ define a fiducial plane. The RVM makes the following assumptions: a zero emission height, the emission is tangent to the local magnetic field, the pulsar's co-rotational speed at the emission altitude is non-relativistic, the emission beams are circular, the magnetic field is well approximated by a static vacuum dipole magnetic field, and the plane of polarisation is parallel to the local magnetic field. 

\section{Method}
\subsection{Folding of Data}
\noindent We used the PPA data from \cite{2017NatAs...1E..29B} obtained on 14 March 2016 in the $340-900$ nm range. We normalised the data to the spin period of the white dwarf by converting the time, which was in Barycentric Julian Day (BJD), to seconds and dividing by the spin period. We then found the minimum PPA of the dataset to define $t_{0}$, which is the starting point used to fold the dataset. We used a standard folding technique to convert to cyclic time. About $5\%$ of the data points deviated from the average curve because of a $180^{\circ}$ ambiguity in the PPA. The folding was found to be affected by the choice of $t_{0}$. The problem was solved by generating a smoothed PPA curve using the Kernel Density Estimation method and a Gaussian kernel e.g. Silverman \cite{silverman2018density}. Depending on the deviation between the smoothed curve and the folded data, we assigned a new convention to the points with a large deviation by shifting these points by $180^{\circ}$ (since there is a $180^{\circ}$ ambiguity in PPA when deriving the RVM). We then binned the data to 30 rotational phase bins.

\subsection{Fixing PPA Discontinuities}
\noindent Using the RVM, Equation \ref{eq:1}, causes the model to be discontinuous since the arctangent function is discontinuous. To make the model continuous, we calculated where $\phi$ made the denominator of Equation \ref{eq:1} go to zero:
\begin{equation} \label{eq:2}
\phi_{\rm{discontinuous}} = \arccos\left( \frac{\tan\zeta}{\tan\alpha}\right). 
\end{equation}
Using this equation, we shift the predicted PPA by $360^{\circ}$ at the points where the model was discontinuous for the different cases to get a smooth, continuous PPA prediction.         
\subsection{Code Verification and Best Fit}
\noindent We used the convention of Everett and Weisberg \cite{2001ApJ...553..341E}, letting $\psi$ increase in the counter-clockwise direction, where we define $\psi' = -\psi$. We then calibrated our model by comparing our best fit  against independent RVM fits to radio pulsar data \citep{2001ApJ...553..341E}, and finding the same results within error margins for $\alpha$ and $\zeta$. The ``best fit'' ($50^{th}$ percentile or median in the posterior distribution of fit parameters) was found using a Markov-Chain Monte Carlo technique \citep{2013PASP..125..306F}. We fit for $\cos\alpha$ and $\cos\zeta$, since we are assuming that $\cos\zeta$ is uniformly distributed.

\section{Results}

\begin{figure}[!h]
\begin{minipage}{18pc}
\includegraphics[width=18pc]{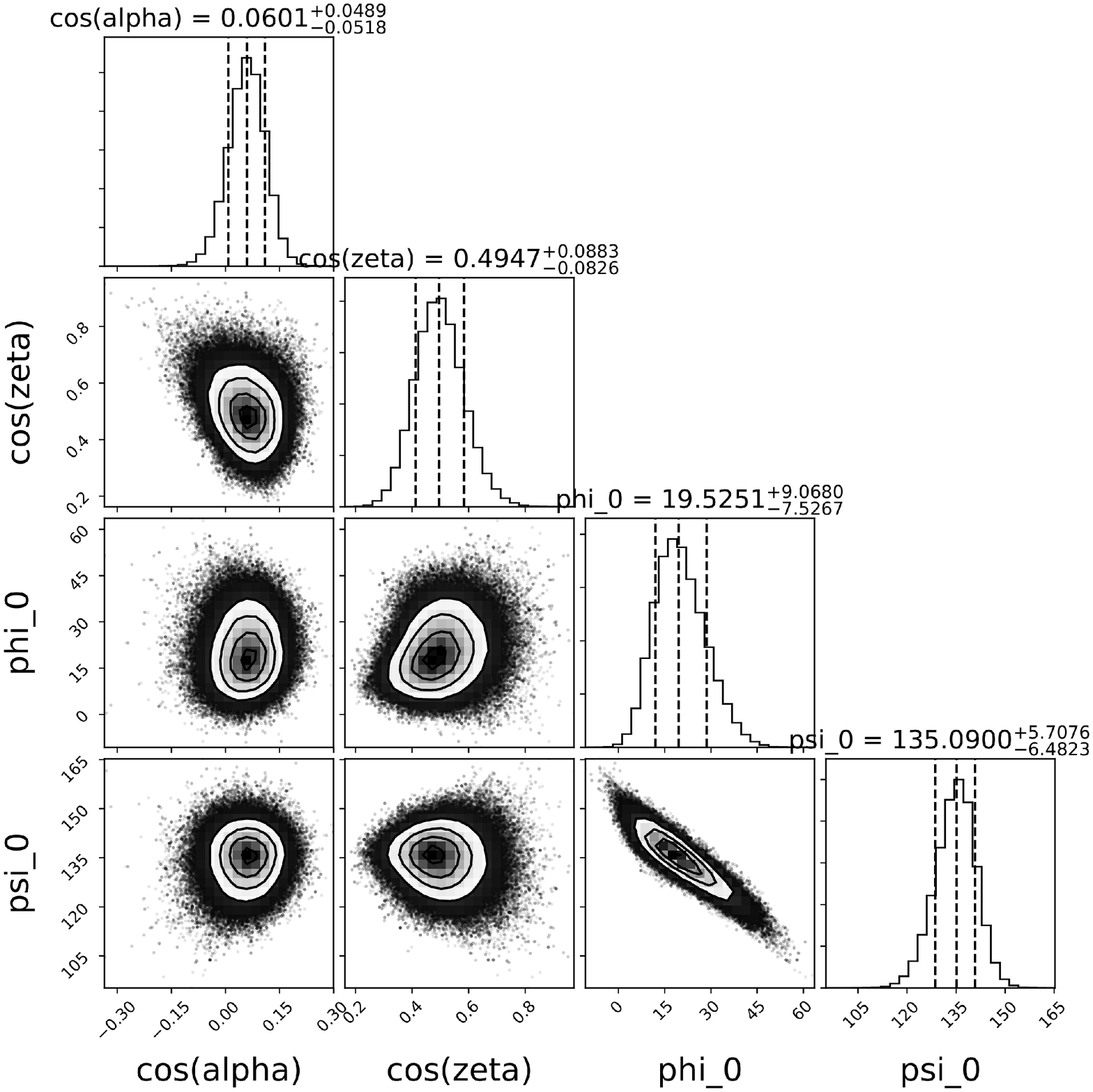}
\label{fig:1} 
\caption{The best fit for the model using $\cos\alpha$ and $\cos\zeta$ The parameter values found were $\cos\alpha=0.060^{+0.049}_{-0.052}$, $\cos\zeta=0.50^{+0.09}_{-0.08}$, $\phi_{0}=(19.5^{+9.1}_{-7.5})^{\circ}$, $\psi_{0}=(135.1^{+5.7}_{-6.5})^{\circ}$. The later 2 are nuisance parameters.}
\end{minipage}\hspace{2pc}%
\begin{minipage}{18pc}
\includegraphics[width=20pc]{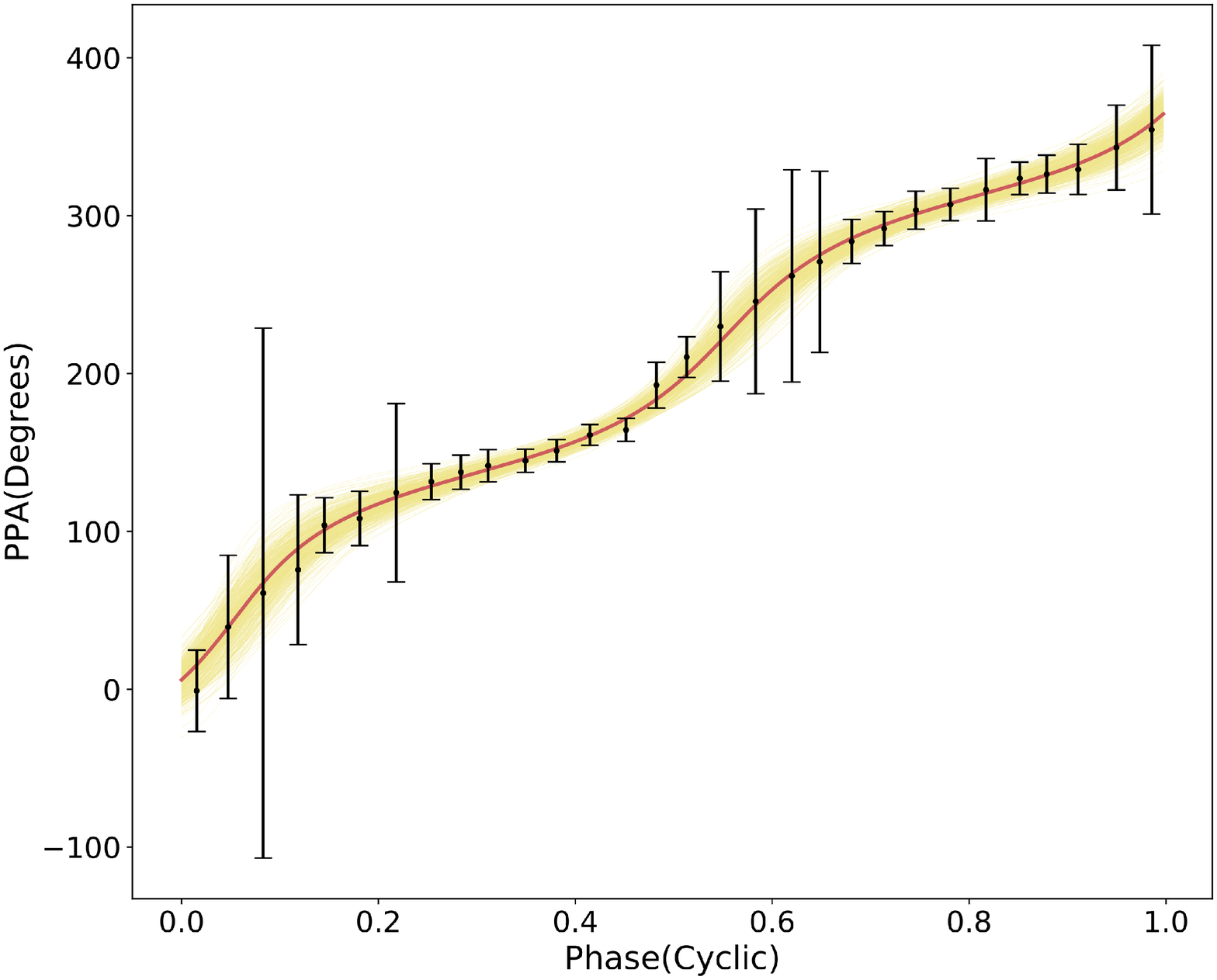} 
\label{fig:2} 
\caption{The best fit for the PPA data using an MCMC technique showing ensemble plots (possible fits).}
\end{minipage} 
\end{figure}

\noindent We found that we can model the white dwarf pulsar using the traditional RVM model. When we found the best fit using the angles directly, it led to disconnected contours in parameter space. We then found the best fit using the cosine of the angles. We found a unique fit for the RVM to the PPA data, namely $\alpha=86.6^{+3.0}_{-2.8}$ and $\zeta=60.4^{+5.3}_{-6.0}$. The large duty cycle leads to relatively small errors on $\alpha$ and $\zeta$. The red curve in Figure \ref{fig:2} shows the best fit and the yellow curves are the ``error band'' derived from the best fit parameters. We also found a best fit with an added likelihood parameter $f$, which compensates for errors underestimated by some fraction $f$.

\section{Discussion}
\noindent We found $\alpha \sim 90^{\circ}$ confirming the expected value from light curve inspection. Since the PPA of the white dwarf could be modelled by the RVM this could mean that the emission of the AR Sco system originates from the magnetic poles of the white dwarf. In the future we want to fit the RVM to PPA data in other energy bands, and for different orbital phases. We will also compare our modelling to the work of Potter and Buckley (in preparation). We furthermore want to incorporate a non-zero emission height in the RVM and apply more detailed emission models to AR Sco. 

\section{Acknowledgements}
\noindent The funding for L.P. was given by the National Astrophysics and Space Science Program (NASSP). This work is based on the research supported wholly in part by the National Research Foundation of South Africa (NRF; Grant Number 99072). The Grantholder acknowledges that opinions, findings and conclusions or recommendations expressed in any publication generated by the NRF supported research is that of the author(s), and that the NRF accepts no liability whatsoever in this regard.

\section*{Appendix: RVM Derivation}
\noindent For an alternative derivation, see Lyutikov \cite{2016arXiv160700777L} using similar techniques. The equation for the magnetic field of a static vacuum magnetic dipole is
\begin{equation}
\mathbf{B} = \dfrac{B_{\rm{p}}}{2r^{3}}\left(2\cos\theta \boldsymbol{\hat{r}} + \sin\theta \boldsymbol{\hat{\theta}} \right), 
\end{equation}
where $B_{\rm{p}}$ is the magnetic field strength at the polar cap, and $r$ and $\theta$ (colatitude) are polar coordinates. Calculating the unit vector for the magnetic field yields
\begin{equation}\label{eq: 5.1}
\mathbf{\hat{B}} = \dfrac{2\cos\theta \boldsymbol{\hat{r}} + \sin\theta \boldsymbol{\hat{\theta}}}{\sqrt{1+3\cos^{2}\theta}}.
\end{equation}
A Cartesian representation of Equation (\ref{eq: 5.1}) is given by 
\begin{equation}
\hat{B} = \sin\gamma\cos\phi\boldsymbol{\hat{x}} + \sin\gamma\sin\phi\boldsymbol{\hat{y}} + \cos\gamma\boldsymbol{\hat{z}} 
\end{equation}
where~ \citep{2018ApJ...854...98W} 
\begin{equation}
\cos\gamma = \dfrac{3\cos^{2}\theta-1}{\A{\theta}}, \quad \sin\gamma = \dfrac{3\cos\theta\sin\theta}{\A{\theta}}.
\end{equation}
Let us define three rotation matrices
\begin{equation}
\Lambda_{\Omega t}= \begin{bmatrix} \cos\Omega t&-\sin\Omega t&0 \\ \sin\Omega t&\cos\Omega t&0 \\0&0&1 \end{bmatrix},
\Lambda_{\alpha}= \begin{bmatrix} \cos\alpha&0&\sin\alpha \\ 0&1&0 \\ -\sin\alpha &0&\cos\alpha \end{bmatrix}, 
\Lambda_{\zeta}= \begin{bmatrix} \cos\zeta&-\sin\zeta&0 \\ \sin\zeta&\cos\zeta&0 \\0&0&1 \end{bmatrix}
\end{equation}
where $\Omega$ is the angular velocity of the pulsar, $\alpha$ is the inclination angle, and $\zeta$ is an arbitrary observer angle. The total rotational matrix is equal to $\Lambda_{\rm tot} = \Lambda_{\zeta}\Lambda_{\Omega t}\Lambda_{\alpha}$ where
\begin{equation}
\Lambda_{\rm{tot}} = \begin{bmatrix}
\cos\Omega t\cos\alpha\sin\zeta-\sin\alpha\cos\zeta& -\sin\Omega t\sin\zeta& \cos\alpha\cos\zeta+\cos\Omega t\sin\alpha\sin\zeta \\ \cos\alpha\sin\Omega t& \cos\Omega t& \sin\alpha\sin\Omega t \\ -\sin\alpha\sin\zeta-\cos\Omega t\cos\alpha\cos\zeta& \cos\zeta\sin\Omega t&\cos\alpha\sin\zeta-\cos\Omega t\sin\alpha\cos\zeta
\end{bmatrix}.
\end{equation}
We apply this rotation matrix to the magnetic field $\mathbf{\hat{B}} $, and choose $\zeta$ such that $(\Lambda_{\rm tot} \cdot \hat{B})_{y} =(\Lambda_{\rm tot} \cdot \hat{B})_{z}=0$. The left-most term yields
\begin{equation} \label{eq: 5.2}
\cos\gamma = -\dfrac{\cos\Omega t\sin\gamma\sin\phi}{\sin\alpha\sin\Omega t} - \dfrac{\cos\alpha\cos\phi\sin\gamma}{\sin\alpha},
\end{equation}
and the $z$-component gives
\begin{equation} \label{eq: 5.3}
\begin{split}
\cos\gamma(\cos\alpha\sin\zeta-\cos\Omega t\sin\alpha\cos\zeta) &- \cos\phi\sin\gamma(\sin\alpha\sin\zeta+\cos\alpha\cos\Omega t\cos\zeta) \\
& + \sin\gamma\cos\zeta\sin\Omega t\sin\phi = 0.  
\end{split}
\end{equation} 
Combing Equations (\ref{eq: 5.2}) and (\ref{eq: 5.3}) and dividing by $\sin\gamma\cos\phi\cos\Omega t\cos\alpha$ yields 
\begin{equation} \label{eq: 5.4}
-\dfrac{\sin\zeta}{\sin\gamma\sin\Omega t}\tan\phi - \dfrac{\cos\alpha\sin\zeta}{\sin\alpha\cos\Omega t} + \dfrac{\cos\Omega t\cos\zeta}{\sin\Omega t\cos\alpha}\tan\phi - \dfrac{\sin\alpha\sin\zeta}{\cos\Omega t\cos\alpha} + \dfrac{\cos\zeta\sin\Omega t}{\cos\Omega t\cos\alpha}\tan\phi = 0.
\end{equation}
Combining the $\tan\phi$ and non-$\tan\phi$ terms elegantly reduces Equation (\ref{eq: 5.4}) to 
\begin{equation} \label{eq: 5.5}
\tan\phi = \dfrac{\sin\zeta\sin\Omega t}{\cos\zeta\sin\alpha-\sin\zeta\cos\alpha\cos\Omega t}.
\end{equation}
We assume that the polarisation is perpendicular to the azimuthal vector $\boldsymbol{e_{\phi}} = (-\sin\phi,\cos\phi,0)$, meaning $\boldsymbol{e_{p}} = \boldsymbol{e_{\phi}}\times\boldsymbol{\hat{n}}$ where $\boldsymbol{\hat{n}}$ is the line-of-sight vector. The azimuthal vector is aligned with $\boldsymbol{\hat{n}}$ by applying the rotation matrix, where the primed coordinates are in the co-rotating frame. Calculating $\boldsymbol{e_{\phi}}^\prime = \Lambda_{\rm tot}  \boldsymbol{e_{\phi}}$ yields
\begin{equation}
\boldsymbol{e_{\phi}}' = \begin{bmatrix}
\sin\phi( \sin\alpha\cos\zeta -\cos\alpha\cos\Omega t\sin\zeta) -\cos\phi\sin\Omega t\sin\zeta \\ \cos\Omega t\cos\phi -\cos\alpha\sin\Omega t\sin\phi \\ \sin\phi( \sin\alpha\sin\zeta +\cos\alpha\cos\Omega t\cos\zeta) +\cos\phi\sin\Omega t\cos\zeta 
\end{bmatrix}.
\end{equation}
The polarisation angle can now be calculated by $\boldsymbol{e_{p}}'=\boldsymbol{e_{\phi}}'\times\boldsymbol{\hat{n}}'$,where $\boldsymbol{\hat{n}}'=(1,0,0)$, which gives $\boldsymbol{e_{p}}'=[0, (\boldsymbol{e_{\phi}}')_{z}, -(\boldsymbol{e_{\phi}}')_{y}]^{T}$. 
The polarisation position angle $\Psi$ can now be calculated using $\tan\Psi=\dfrac{(\boldsymbol{e_{p}}')_{y}}{(\boldsymbol{e_{p}}')_{z}}$. Upon simplification we find
\begin{equation} \label{eq: 5.6}
\tan\Psi = \dfrac{\tan\phi(\sin\alpha\sin\zeta+\cos\Omega t\cos\alpha\cos\zeta)+\cos\zeta\sin\Omega t}{\cos\alpha\sin\Omega t\tan\phi -\cos\Omega t}.
\end{equation}
Substituting Equation (\ref{eq: 5.5}) into Equation (\ref{eq: 5.6}) and simplifying yields
\begin{equation}
\tan\Psi = \dfrac{\sin\alpha\sin\Omega t}{\sin\zeta\cos\alpha-\sin\alpha\cos\zeta\cos\Omega t}.
\end{equation}

\bibliographystyle{JHEP}
\bibliography{bibfile}

\end{document}